\documentclass[aps,prb,10pt,amsmath,amssymb,twocolumn,longbibliography]{revtex4-2}
\usepackage[dvipdfmx]{hyperref}
\usepackage{graphicx}
\usepackage{mathtools}
\usepackage{physics}

\def\be#1\ee{\begin{align}#1\end{align}}

\begin{document}

\title{Polarization jumps by breaking symmetries of two-dimensional Weyl semimetals}

\author{Hiroki Yoshida}
\affiliation{Department of Physics, Tokyo Institute of Technology, 2--12--1 Ookayama, Meguroku, Tokyo 152--8551, Japan}
\author{Tiantian Zhang}
\affiliation{Department of Physics, Tokyo Institute of Technology, 2--12--1 Ookayama, Meguroku, Tokyo 152--8551, Japan}
\author{Shuichi Murakami}
\affiliation{Department of Physics, Tokyo Institute of Technology, 2--12--1 Ookayama, Meguroku, Tokyo 152--8551, Japan}

\date{\today}

\begin{abstract}
  The electric polarization as a bulk quantity is described by the modern theory of polarization in insulating systems and cannot be defined in conducting systems. Upon a gradual change of a parameter in the system, the polarization always varies smoothly as long as the gap remains open. In this paper, we focus on the two-dimensional Weyl semimetal, which hosts Weyl nodes protected by symmetries, and study the behavior of the polarization when a symmetry-breaking term $M$ is introduced and a gap opens. We show that there can be a jump between $M\to0^+$ and $M\to0^-$ limits. We find that the jump is universally described by the ``Weyl dipole" representing how the Weyl nodes with monopole charges are displaced in the reciprocal space. Our result is applicable to general two-dimensional Weyl semimetals.
\end{abstract}

\maketitle

\section{Introduction}
\label{sec:introduction}

The electric polarization is a fundamental quantity when we deal with ferroelectric materials. Despite its importance, the definition of the electric polarization in periodic systems remained elusive for a long time and caused a lot of debates because the expectation value of the position of an electron is ill-defined in periodic systems. The problem was solved by the modern theory of polarization, which tells us that the change of polarization is a well-defined quantity and the polarization itself is multi-valued~\cite{Resta1992,King-Smith1993,Vanderbilt1993}. Namely, the polarization vector $\vb{P}$ is defined modulo the ``quantum of polarization" $\frac{e}{\Omega}\vb{a}$, where $\vb{a}$ is the lattice vector, $-e$ is an electron charge $(e>0)$, and $\Omega$ is the volume of the unit cell. The basic understanding is given by this modern theory of polarization and now it is commonly used to numerically calculate electric polarizations of various insulating systems.

On the other hand, topological states of matter have attracted much interest recently, and theoretical and experimental investigations of topological insulators~\cite{Kane2005a,Kane2005b,Fu2007,Konig2010,Bernevig2006} and semimetals~\cite{Wan2011,Huang2015,Lv2015,Xu2015,Wang2012,Young2012,Young2015} have been active. In particular, a class of materials called Weyl semimetal has been studied intensively because of its unique topological surface states~\cite{Wan2011}, magnetoelectric responses~\cite{Vazifeh2013,Chan2017}, and transport properties~\cite{Gorbar2018}. In three-dimensional Weyl semimetals, there exist gap-closing points called Weyl nodes in the reciprocal space, which carry monopole charges of $\pm Q$ ($Q$: integer). These points cannot be gapped unless Weyl nodes with opposite monopole charges meet and are annihilated. This leads to the stability of Weyl semimetals.

In this paper, we study two-dimensional Weyl semimetals whose Weyl nodes are protected by symmetries such as $\mathcal{PT}$-symmetry ($\mathcal{P}$: inversion, $\mathcal{T}$: time-reversal). We introduce a real symmetry-breaking term $M$ and we assume that the system becomes gapped when $M\neq0$. Then, the modern theory of polarization can be applied after breaking the symmetries, i.e. when $M\neq0$. We show that in the limit where the symmetries are restored and the system becomes conducting, there can be a difference in the value of the electric polarization between $M\to0^+$ and $M\to0^-$, which leads to a jump of polarization. Furthermore, we show that this jump of polarization is closely related with monopole charges of Weyl nodes and can be universally described by the Weyl dipole, which we introduce in this paper.

We note that our focus is not on the behavior of the electric polarization in conducting systems as discussed in several studies~\cite{Mahon2019,Mahon2021}. Although we investigate the conducting limit, our interest is still in the insulating systems, where the modern theory of polarization can be applied. Also, we note that the polarization jump we discuss here occurs in a single material unlike the one at the interface of two insulators~\cite{Ohtomo2004}.

This paper is organized as follows. In Sec.~\ref{sec:example}, we introduce a two-dimensional two-band tight-binding model of a hexagonal lattice with anisotropic hoppings. We numerically see the existence of the polarization jump across a $\mathcal{PT}$-symmetric Weyl semimetal phase as a typical example. The analysis is extended to general multiband systems in Sec.~\ref{sec:general}. We see that the jump of electric polarization can be described in an amazingly simple form by the Weyl dipole, which represents how the Weyl nodes of opposite monopole charges are displaced in $\vb{k}$-space. We conclude this paper in Sec.~\ref{sec:conclusion}. Throughout this paper, we study noninteracting electonic systems.

\section{Example: two-dimensional tight-binding model on the hexagonal lattice}
\label{sec:example}

\subsection{Model}
\label{subsec:model}

\begin{figure}[t]
  \begin{center}
    \includegraphics[width=\columnwidth]{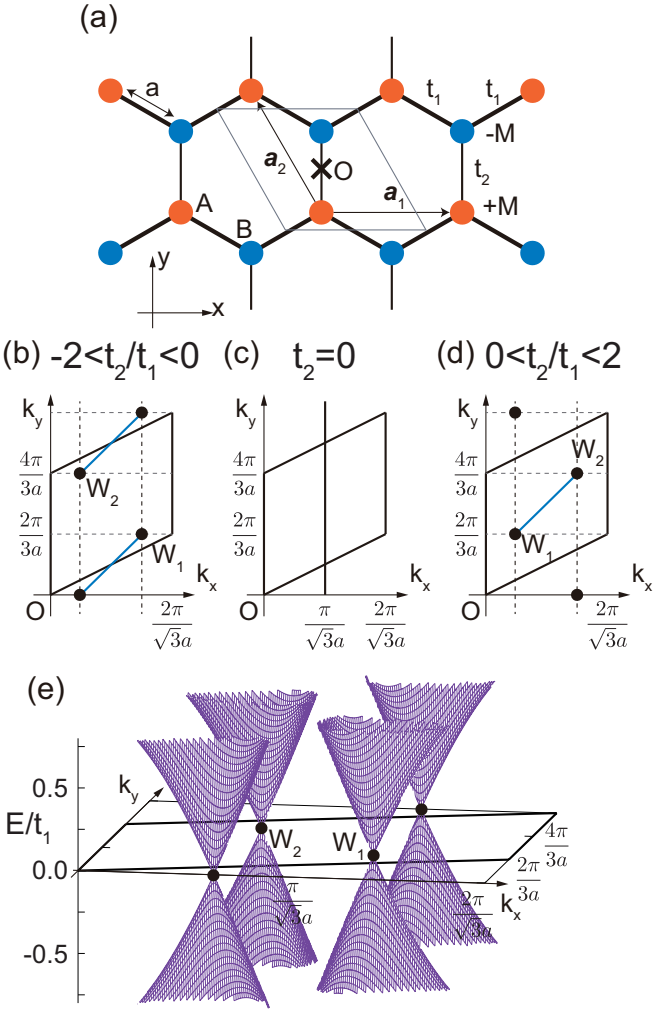}
    \caption{Two-dimensional tight-binding model having Weyl nodes in the $\mathcal{PT}$-symmetric limit. (a) The model on the hexagonal lattice with anisotropic hoppings and on-site potentials $+M$ and $-M$ for A (red) and B (blue) sublattices, respectively. We take the origin O to be at the midpoint of a nearest-neighbor bond and the unit cell as a gray parallelogram. (b-d) The corresponding unit cell of the reciprocal space in the shape of the parallelogram. The Weyl nodes ($W_1$ and $W_2$) for $M=0$ are shown for (b) $-2<t_2/t_1<0$ and (d) $0<t_2/t_1<2$. The $2\pi$ jump in $\arg\qty(f^*(\vb{k}))$ restricted within $-\pi<\arg(f^*)\leq\pi$ is also shown in (b) and (d) by blue lines connecting the Weyl nodes. When (c) $t_2=0$, the gap closes along $k_x=\frac{\pi}{\sqrt{3}a}$ indicated by the black line. (e) Energy bands corresponding to $M=0$ and $-2<t_2/t_1<0$ in the vicinity of the Fermi energy.}
    \label{fig:BN_lattice}
  \end{center}
\end{figure}

We first introduce a two-dimensional tight-binding model on the hexagonal lattice as in Fig.~\ref{fig:BN_lattice}(a). In the figure, red and blue circles represent sites in A and B sublattices with on-site potentials $+M$ and $-M$, respectively. Among the three hopping amplitudes to the nearest neighbor sites, let $t_2$ and $t_1$ denote the amplitude in the $y$-direction and those in other directions, respectively, where $t_1(>0)$ and $t_2$ are real. Here, $\vb{a}_1=(\sqrt{3}a,\ 0)$ and $\vb{a}_2=\qty(-\frac{\sqrt{3}a}{2},\ \frac{3a}{2})$ are the primitive lattice vectors, where $a$ is the lattice constant. Then, the spinless tight-binding Hamiltonian of this system is
\be
  \hat{H} &= \sum_{\langle i,j\rangle}t_{ij}\hat{c}_i^{\dagger}\hat{c}_j+\sum_i m_i\hat{c}_i^{\dagger}\hat{c}_i,\label{eq:hamiltonian}
\ee
where $\hat{c}_i\ (\hat{c}_i^{\dagger})$ annihilates (creates) an electron at site $i$, and $\langle i,j \rangle$ denotes a pair of nearest-neighbor sites. Here, the first term represents nearest neighbor hoppings with $t_{ij}=t_2$ for the $y$-direction and $t_{ij}=t_1$ otherwise, and the second term represents an on-site potential $m_i=+M$ and $-M$ for A and B sublattices, respectively. When $M=0$, the system is $\mathcal{PT}$-symmetric. A non-zero value of $M$ breaks the inversion symmetry and preserves the time-reversal symmetry. Within the basis of sublattices A and B, the $\vb{k}$-dependent Bloch Hamiltonian is given by
\be
  H(\vb{k}) &= \mqty(M&f^*(\vb{k})\\f(\vb{k})&-M),\label{eq:hamiltonian_matrix}\\
  f(\vb{k}) &\coloneqq 2t_1\cos\qty(\frac{\sqrt{3}a}{2}k_x)e^{i\frac{a}{2}k_y}+t_2e^{-iak_y}.\label{eq:f}
\ee
We take the unit cell of the reciprocal space of this system to be a parallelogram spanned by reciprocal lattice vectors $\vb{b}_1=\qty(\frac{2\pi}{\sqrt{3}a},\ \frac{2\pi}{3a})$ and $\vb{b}_2=\qty(0,\ \frac{4\pi}{3a})$ as shown in Figs.~\ref{fig:BN_lattice}(b), (c) and (d). Energies of this system are given by $E_{\pm}=\pm\sqrt{M^2+|f|^2}$. We set the system to be half-filled and the Fermi energy at $E_F=0$.

This system becomes conducting if and only if $M=0$ and $f(\vb{k})=0$ at some $\vb{k}$ points. In this case, the band structure has two Weyl nodes located at
\be
 W_1 &: \qty(\frac{2}{\sqrt{3}a}\cos^{-1}\qty(\frac{t_2}{2t_1}),\ \frac{2\pi}{3a}),\label{eq:W1}\\
 W_2 &: \qty(\frac{2}{\sqrt{3}a}\cos^{-1}\qty(-\frac{t_2}{2t_1}),\ \frac{4\pi}{3a}),\label{eq:W2}
\ee
which are related by the mirror symmetry with respect to the $y$ axis. The energies in the vicinity of the Fermi energy and Weyl nodes are shown in Fig.~\ref{fig:BN_lattice}(e) for $M=0$ and $-2<t_2/t_1<0$ as an example. These Weyl nodes exist if $|t_2/t_1|\leq2$ and hence, the electric polarization of this system is not defined when parameters are in the region $M=0$ and $|t_2/t_1|\leq2$. These Weyl nodes move parallel to the $k_x$ axis when we change $t_2/t_1$. When $t_2/t_1$ is negative, the $k_x$ components of the Weyl nodes satisfy $W_{2,x}<W_{1,x}$ as in Fig.~\ref{fig:BN_lattice}(b) and when $t_2/t_1$ is positive, $W_{2,x}>W_{1,x}$ as in Fig.~\ref{fig:BN_lattice}(d). The case with $M=0$ and $t_2=0$ is an exception, where the band gap closes along the line $k_x=\frac{\pi}{\sqrt{3}a}$ (Fig.~\ref{fig:BN_lattice}(c)).

We next calculate the electric polarization of this system. According to the modern theory of polarization, the electronic contribution to the electric polarization is expressed in terms of the Berry phase of Bloch states. For two-dimensional systems, the contribution of electrons to the electric polarization $\vb{P}^e$ in the $\alpha$-direction is given by
\be
  P^e_{\alpha} = \frac{-ie}{(2\pi)^2}\int_{\mathrm{BZ}}\mathrm{d}^2k \sum_{n}^{occ}\bra{u_n(\vb{k})}\pdv{}{k_{\alpha}}\ket{u_n(\vb{k})}\label{eq:polarization},
\ee
where the integral is over the Brillouin zone, the sum is taken over all the occupied bands, and $\ket{u_n(\vb{k})}$ is the cell periodic part of the Bloch state of the $n$-th energy level. This expression is valid under the wavevector periodic gauge~\cite{King-Smith1993}. Under this gauge, energy eigenfunctions $u_{\vb{r},n}(\vb{k})$ satisfy the condition $u_{\vb{r},n}(\vb{k})=e^{i\vb{b}\cdot\vb{r}}u_{\vb{r},n}(\vb{k}+\vb{b})$, where $\vb{b}$ is the reciprocal lattice vector. This gauge condition depends on the choice of the origin. For convenience, we take the origin O of the model to be at the midpoint between the A and B sites constituting the unit cell (Fig.~\ref{fig:BN_lattice}(a)), so that the model at $M=0$ preserves inversion symmetry with respect to the origin.
To make the electric polarization vector $\vb{P}$ well-defined, we need to incorporate the ionic contributions $\vb{P}^{ion}$ and make the system charge neutral. The ionic polarization is the sum of products of charges and positions of all the ions in the unit cell. For convenience, we assume that the ions are distributed in an inversion-symmetric manner, which leads to $\vb{P}^{ion}=\vb{0}$. Then, Eq.~\eqref{eq:polarization} gives the electric polarization of the system. Since we have set $E_F=0$, we need one energy eigenstate corresponding to the lower energy $E_-=-\sqrt{M^2+\abs{f}^2}$. The choice of the origin affects the choice of the gauge via the relation $u_{\vb{r},n}(\vb{k})=e^{i\vb{b}\cdot\vb{r}}u_{\vb{r},n}(\vb{k}+\vb{b})$. Here, the gauge choice of the energy eigenstates needs special attention since it is not possible to take a gauge which is analytical in the entire parameter space. Depending on the sign of the parameter $M$, we take
\be
  \ket{u^+(\vb{k})}&=c_{+}\mqty(-\frac{f^*}{M+\sqrt{M^2+|f|^2}}\\1)e^{-i\frac{a}{2}k_y}\label{eq:eigvec+}
\ee
for $M>0$ and
\be
  \ket{u^-(\vb{k})}&=c_{-}\mqty(1\\\frac{f}{M-\sqrt{M^2+|f|^2}})e^{i\frac{a}{2}k_y}\label{eq:eigvec-}
\ee
for $M<0$, where the coefficients $c_{+}$ and $c_{-}$ are the normalization factors. We chose the different wavefunctions for $M>0$ and $M<0$ in order to avoid divergence in their components. By choosing gauges in such a manner, the Berry connection can be written as
\be
  &i\bra{u^{\pm}}\pdv{}{k_{\alpha}}\ket{u^{\pm}}\nonumber\\
  &\quad=\frac{\mp |f|^2}{|f|^2+(M\pm\sqrt{M^2+|f|^2})^2}\pdv{}{k_\alpha}\arg(f^*)\pm\frac{a}{2}\delta_{\alpha,y}\label{eq:BerryConnection},
\ee
where the upper and lower signs correspond to the cases with $M>0$ and $M<0$, respectively. By integrating these quantities over the Brillouin zone, we get the $\alpha$ component of the electric polarization vector $\vb{P}$.

\subsection{Polarization and edge charge density}
\label{subsec:polarization}

\begin{figure}[t]
  \begin{center}
    \includegraphics[width=\columnwidth]{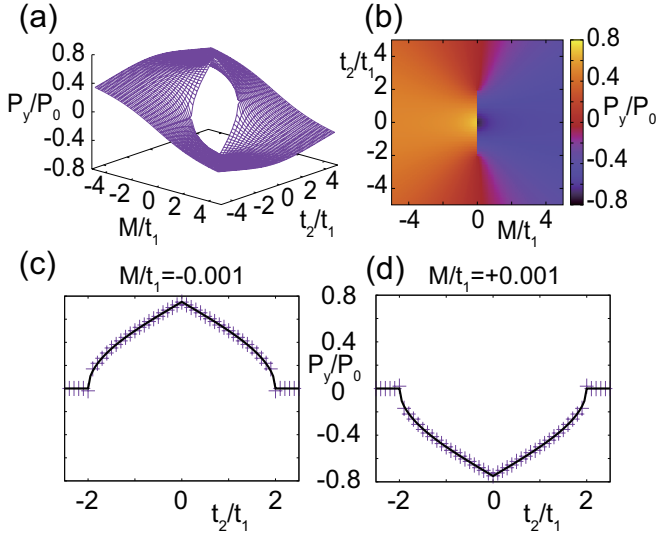}
    \caption{The $y$ component of the total polarization vector  $\vb{P}$. (a) Three-dimensional plot and (b) two-dimensional color plot of $P_y$ in the parameter space. We put $P_0=\frac{ae}{\Omega}$, where $\Omega$ is the area of the unit cell. (c) (d) $P_y$ at (c) $M/t_1=-0.001$ and at (d) $M/t_1=0.001$. Results of numerical calculations using Eq.~\eqref{eq:BerryConnection} are plotted by markers and the black lines show plots of Eq.~\eqref{eq:AnalyticalP}. There exists the polarization jump across $M=0$ when $-2<t_2/t_1<2$, where the system is a Weyl semimetal. When $|t_2/t_1|>2$, the polarization is zero at $M=0$ because of the inversion symmetry.}
    \label{fig:BN_P}
  \end{center}
\end{figure}

Next, we numerically calculate the electric polarization $\vb{P}$ of this system by using Eqs.~\eqref{eq:polarization} and \eqref{eq:BerryConnection}. From the mirror symmetry of the system with respect to the $y$ axis, $P_x$ is trivially zero as a function of $M/t_1$ and $t_2/t_1$, as is confirmed directly from Eq.~\eqref{eq:BerryConnection}. Hence, we only focus on the $y$ component of the polarization. By calculating $P_y$ for various parameter values of $M/t_1$ and $t_2/t_1$, the electric polarization in the parameter space is plotted in Figs.~\ref{fig:BN_P}(a) and (b). In this system, the polarization is defined in terms of modulo the quantum of polarization $\frac{e}{\Omega}\vb{a}_{1,2}$, where $\Omega = \frac{3\sqrt{3}a^2}{2}$ is the area of the unit cell. We see that the polarization changes smoothly and continuously in regions where the system is insulating. Meanwhile, the system is a Weyl semimetal when $|t_2/t_1|<2$ and $M=0$, and the polarization is not defined. When $|t_2/t_1|<2$, the limit value of $P_y$ taking $M$ to zero from the positive side, $\lim_{M\to0^{+}}P_y$, and that from the negative side, $\lim_{M\to0^{-}}P_y$, give different values.

To study the behavior of the polarization in the limits $M\to0^{\pm}$, we note that the $y$ component of the Berry connection can be simplified in the limit $M\to0^{\pm}$ as
\be
  \lim_{M\to0^{\pm}}i\bra{u^{\pm}}\pdv{}{k_{y}}\ket{u^{\pm}}&=\mp \frac{1}{2}\pdv{}{k_{y}}\arg(f^*)\pm \frac{a}{2}.\label{eq:BerryConnectionM0lim}
\ee
This means that the polarization in the $y$-direction of the system can be written as a change of $\arg(f^*)$ across the Brillouin zone. We then notice that $f^*$ vanishes at the Weyl nodes $W_1$ and $W_2$, and that $\arg(f^*)$ winds around these Weyl nodes. Hence, there has to be a jump of $\arg(f^*)$ across a curve in reciprocal space connecting two Weyl nodes if we set $-\pi<\arg(f^*)\leq\pi$. Examples of the position of the jump are shown by the blue lines in Figs.~\ref{fig:BN_lattice}(b) and (d) for negative and positive values of $t_2/t_1$, respectively. By crossing these lines, $\arg(f^*)$ changes by $2\pi$, and hence, the region between Weyl nodes projected to the $k_x$ axis needs special care when we integrate over $k_x$. For the limit $M\to0^{\pm}$, we can find that
\be
  \int_{k_y^0}^{k_y^0+\frac{4\pi}{3a}}\mathrm{d}k_y\lim_{M\to0^{\pm}}i\bra{u^{\pm}}\pdv{}{k_{y}}\ket{u^{\pm}}&=\left\{\begin{array}{l}0\quad(k_x\in W)\\\pm\pi\quad(\mathrm{otherwise}),\end{array}\right.
\ee
where $W$ is an open interval $(W_{2,x},W_{1,x})$ for $t_2/t_1<0$ and $(W_{1,x},W_{2,x})$ for $t_2/t_1>0$, respectively, and $k_y^0\coloneqq \frac{1}{\sqrt{3}}k_x$. Then, using positions of two Weyl nodes Eqs.~\eqref{eq:W1} and \eqref{eq:W2}, the $y$ components of the polarization in the $M\to0^{\pm}$ limits for $|t_2/t_1|<2$ are given by
\be
  P_y^{\pm}&=\frac{-e}{(2\pi)^2}\qty(\frac{2\pi}{\sqrt{3}a}-\abs{W_{2,x}-W_{1,x}})\times\qty(\pm \pi) \nonumber\\
  &=\mp\qty(\frac{3}{4}-\frac{3}{4\pi}\abs{\cos^{-1}\qty(-\frac{t_2}{2t_1})-\cos^{-1}\qty(\frac{t_2}{2t_1})})P_0\label{eq:AnalyticalP},
\ee
where $P_0\coloneqq\frac{ae}{\Omega}$. On the other hand, for $|t_2/t_1|>2$, $P_y^{\pm}=0$ holds, as expected from the inversion symmetry at $M=0$. We plot these analytical results in Figs.~\ref{fig:BN_P}(c) and (d) at $M\to0^{\mp}$, respectively. The results of numerical integration of Eq.~\eqref{eq:BerryConnection} over the Brillouin zone at $M/t_1=\mp0.001$ are also plotted. We can see that although the numerical calculation in the small $M$ region is difficult due to the small band gap size, numerical and analytical results agree with each other.

In the Weyl semimetal phase ($M=0,|t_2/t_1|<2$), the bands have two Weyl nodes protected by the nontrivial $\pi$ Berry phase around each Weyl node, and these Weyl nodes are distant from each other in $\vb{k}$-space (see Figs.~\ref{fig:BN_lattice}(b) and (d)). Nonetheless, the case with $M=0,t_2=0$ is an exception. At $M=0$ and $t_2=0$, the system is one-dimensional and the band structure is independent of $k_y$; it makes the two gap closing points extend along the $k_y$-direction, and allows them to meet. In the resulting band structure, the gap closes along $k_x=\frac{\pi}{\sqrt{3}a}$ (see Fig.~\ref{fig:BN_lattice}(c)), and the system is equivalent to the one-dimensional chain with nearest-neighbor hopping $t_1$.
Therefore, the $M$ term gives a staggered potential to the one-dimensional chain, and the polarizations for $M=0^+$ and $M=0^-$ are different by $\Delta\vb{P}=\frac{e}{2\Omega}\vb{a}_1=\qty(\frac{\sqrt{3}}{2}P_0,0)\equiv\qty(0,-\frac{3}{2}P_0)\ \qty(\mathrm{mod}\ \frac{e}{\Omega}\vb{a}_{1,2})$, in agreement with our result showing $P^{\pm}_y=\mp\frac{3}{4}P_0$.

\begin{figure}[t]
  \begin{center}
    \includegraphics[width=\columnwidth]{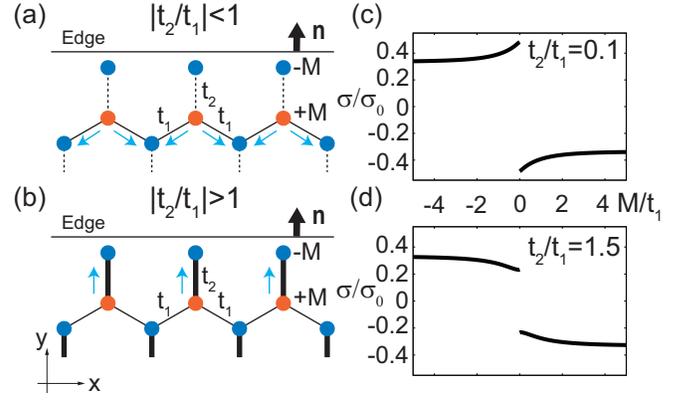}
    \caption{Edge charge densities of the model. (a) (b) The edge of the system with (a) $|t_2/t_1|<1$ and (b) $|t_2/t_1|>1$. Electrons tend to move in the directions of blue arrows when $M/t_1$ is increased. (c) and (d) show the accumulated edge charge density as a function of $M/t_1$ with $t_2/t_1=0.1$ and $1.5$, respectively corresponding to the cases of (a) and (b). $\sigma_0\coloneqq\frac{e}{\sqrt{3}a}$ is the unit of the line charge density.}
    \label{fig:BN_surface}
  \end{center}
\end{figure}

When we make an edge by cutting the system as in Figs.~\ref{fig:BN_surface}(a) and (b), the edge charge density can be obtained from the polarization calculated above as $\sigma = \vb{P}\cdot\vb{n}$, where $\vb{n}=(0,1)$ is the unit vector normal to the edge. We plot $\sigma$ as a function of $M/t_1$ at $t_2/t_1=0.1$ and $1.5$ in Figs.~\ref{fig:BN_surface}(c) and (d), respectively. We can see that the rate of change of the edge charge density by increasing $M/t_1$ has opposite signs between the cases with $|t_2/t_1|<1$ and $|t_2/t_1|>1$. This can be qualitatively understood by considering the motion of electrons by increasing the value of $M/t_1$. When $|t_2/t_1|<1$, electrons at the A sublattice are transferred to the B sublattice by hoppings $t_1$ and electrons tend to move in the directions of blue arrows in Fig.~\ref{fig:BN_surface}(a), leading an increase in the edge charge density. On the other hand, when $|t_2/t_1|>1$, electrons tend to move in the directions of blue arrows in Fig.~\ref{fig:BN_surface}(b) and it leads to the decrease in the edge charge density.

\section{Polarization jump in general systems across a Weyl semimetal phase}
\label{sec:general}

The jump of the electric polarization in the previous section is essentially caused by the $2\pi$ phase winding of $f(\vb{k})$ around the Weyl nodes. Although details of the jump are dependent on the form of $f(\vb{k})$ for the system under consideration, the mechanism is the same for all two-band models described by the Hamiltonian in Eq.~\eqref{eq:hamiltonian_matrix}. In this section, we consider a general two-dimensional $N$-band spinless system in a Weyl semimetal phase. We then introduce a parameter $M$ which breaks the symmetry protecting the Weyl nodes and opens a gap. Then, we show that the jump of the electric polarization across $M=0$ is similarly described to that in the previous section.

\subsection{General theory}
\label{subsec:GeneralTheory}

Consider a general two-dimensional $N$-band model. We assume that the system is in the Weyl semimetal phase where the band gap closes at two Weyl nodes protected by some symmetries. Next, as the example shown in the previous section, we introduce a symmetry-breaking real parameter $M$ to this Hamiltonian. Let us further suppose that the system is insulating when $M\neq0$, because a gap opens at Weyl nodes by broken symmetries. Let $H^{(M)}(\vb{k})$ be the Hamiltonian at the parameter value $M$. Then, the jump of the polarization vector $\Delta \vb{P}$ across $M=0$ along one of the reciprocal vector $\vb{b}_{\beta}$ is
\be
  \Delta P_{\beta}&\coloneqq P_{\beta}^+-P_{\beta}^-\nonumber\\
  &=\frac{-ie}{(2\pi)^2}\int_{\mathrm{BZ}}\mathrm{d}^2k \sum_n^{\mathrm{occ}}\left(\lim_{M\to0^+}\bra{u_n^{(M)}}\pdv{}{k_{\beta}}\ket{u_n^{(M)}}\right. \nonumber\\
  &\qquad\left. -\lim_{M\to0^-}\bra{u_n^{(M)}}\pdv{}{k_{\beta}}\ket{u_n^{(M)}}\right)\label{eq:deltaP},
\ee
where $\ket{u_n^{(M)}}$ is the eigenstate of the $n$-th band of the $M$-dependent Hamiltonian $H^{(M)}(\vb{k})$. Here, the choice of the origin is fixed regardless of the parameter values. Then the ionic polarization does not have a jump at $M=0$ and has no contribution to Eq.~\eqref{eq:deltaP}. The polarization jump across $M=0$ is now proportional to the difference of the Berry phases.

Except for the $\vb{k}$-points where the gap closes at $M=0$, the subspace spanned by the occupied states at a fixed value of $\vb{k}$ changes analytically across $M=0$, and there is no singularity. Then, Eq.~\eqref{eq:deltaP} can be rewritten in terms of the single eigenstate of the top of the valence band near the $\vb{k}$-points where the gap closes at $M=0$. At this point, the expression of the polarization jump is simplified and the Berry phase of only the single band is needed. Therefore, as we explain in the following, we can reduce the problem into a $2\times 2$ effective Hamiltonian, which is easier to handle analytically.

Let $\ket{u_n(\vb{k})}\ (n=1,2,\cdots,N)$ be eigenstates at $M=0$ arranged in an increasing order of the energy, $E_1\leq E_2\leq\cdots\leq E_N$ and $N'$ denote the number of occupied bands when $M\neq0$. Then, in order to calculate the polarization jump across $M=0$, we only need to consider the effective $2\times 2$ Hamiltonian
\be
  &H_{\mathrm{eff}}^{(M)}(\vb{k})\nonumber\\
  &\coloneqq \mqty(\bra{u_{N'+1}}H^{(M)}(\vb{k})\ket{u_{N'+1}} & \bra{u_{N'+1}}H^{(M)}(\vb{k})\ket{u_{N'}}\\
  \bra{u_{N'}}H^{(M)}(\vb{k})\ket{u_{N'+1}} & \bra{u_{N'}}H^{(M)}(\vb{k})\ket{u_{N'}})\label{eq:EffectiveHamiltonian}
\ee
for the two bands $n=N',N'+1$, which constitute the Weyl nodes at $M=0$, and focus on the $\vb{k}$-points where the gap closes at $M=0$. This effective Hamiltonian describes the Weyl semimetal phase at $M=0$, inheriting the symmetry of the original Hamiltonian. Since we got a $2\times2$ matrix, calculations of the Berry connection and the jump of polarization are parallel to those in the previous section. We can write the jump of polarization in the $\beta$-direction across the Weyl semimetal phase as
\be
  \Delta P_{\beta} &= \frac{-e}{(2\pi)^2}\int_0^{b_{\alpha}}\mathrm{d} k_{\alpha}\, \Delta \phi_{\beta}(k_{\alpha})\sin\theta,\label{eq:EffectiveDP}
\ee
where we take $\left\{\vb{b}_{\alpha},\vb{b}_{\beta}\right\}$ be the set of primitive lattice vectors, $k_{\alpha}$ is the wavevector component along $\vb{b}_{\alpha}$ direction, $\theta$ is an angle between $\vb{b}_{\alpha}$ and $\vb{b}_{\beta}$, and $b_{\alpha}=\abs{\vb{b}_{\alpha}}$. Here,
\be
  \Delta\phi_{\beta}(k_{\alpha}) &\coloneqq \phi^+_{\beta}(k_{\alpha})-\phi^-_{\beta}(k_{\alpha})
\ee
is the difference of Berry phases in the $\beta$-direction:
\be
  \phi_{\beta}^{\pm}(k_{\alpha}) &\coloneqq \lim_{M\to0^{\pm}}i\int_0^{b_{\beta}}\mathrm{d} k_{\beta} \bra{u_0^{(M)}}\pdv{}{k_{\beta}}\ket{u_0^{(M)}},
\ee
where $\ket{u_0^{(M)}}$ is the eigenstate corresponding to the lower energy of the effective Hamiltonian, $k_{\beta}$ is the component of $\vb{k}$ along $\vb{b}_{\beta}$, and $b_{\beta}=\abs{\vb{b}_{\beta}}$. This Berry phase difference $\Delta\phi_{\beta}(k_{\alpha})$ can jump as we change $k_{\alpha}$ across the Weyl nodes.

\begin{figure}[t]
  \begin{center}
    \includegraphics[width=0.8\columnwidth]{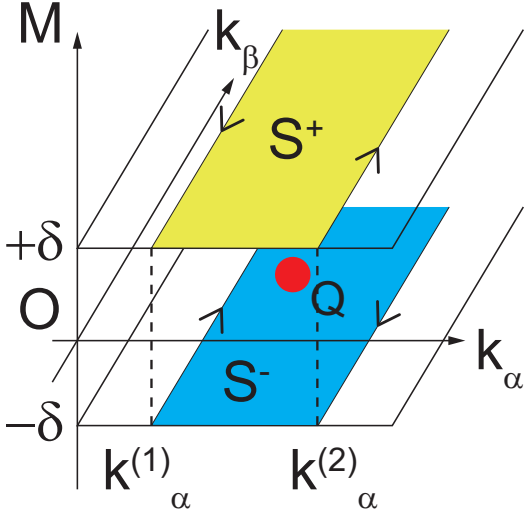}
    \caption{A (higher-order) Weyl node with monopole charge $Q$ in three-dimensional $(k_{\alpha},k_{\beta},M)$ space. Two planes $S^{\pm}$ indicate the area of integration in Eq.~\eqref{eq:diffDeltaphi}.}
    \label{fig:WeylNode}
  \end{center}
\end{figure}

A key step of our calculation is to regard this effective Hamiltonian $H_{\mathrm{eff}}^{(M)}(\vb{k})$ as a Hamiltonian defined in the three-dimensional $(k_x,k_y,M)$ space. Then the two-dimensional Weyl nodes $(W_{j,x},W_{j,y})\ (j=1,2,\cdots)$ can be considered as three-dimensional Weyl nodes $(W_{j,x},W_{j,y},M=0)$, having monopole charges $\pm 1$.
By using this topological nature of Weyl nodes, general understandings of the jump of the polarization can be given. To calculate this jump, let us consider the case in Fig.~\ref{fig:WeylNode}, where a Weyl node with a monopole charge $Q$ lies between $k_{\alpha}=k_{\alpha}^{(1)}$ and $k_{\alpha}^{(2)}$ in the $(\vb{k},M)$ space (Here, this Weyl node can be of higher order with $|Q|\geq2$). Then the difference of $\Delta \phi_{\beta}(k_{\alpha})$ between these two values of $k_{\alpha}$ can be evaluated as
\be
  &\Delta\phi_{\beta}\qty(k_{\alpha}^{(2)})-\Delta\phi_{\beta}\qty(k_{\alpha}^{(1)})\nonumber\\
  &=\qty(\phi_{\beta}^+\qty(k_{\alpha}^{(2)})-\phi_{\beta}^+\qty(k_{\alpha}^{(1)}))-\qty(\phi_{\beta}^-\qty(k_{\alpha}^{(2)})-\phi_{\beta}^-\qty(k_{\alpha}^{(1)}))\nonumber\\
  &=\iint_{S^+}\mathrm{d}\vb{S}\cdot\vb{B}(\vb{k},M)-\iint_{S^-}\mathrm{d}\vb{S}\cdot\vb{B}(\vb{k},M)\nonumber\\
  &=2\pi Q\label{eq:diffDeltaphi},
\ee
in the limit $\delta\to0$, where the surface integrals are evaluated over surfaces $S^{\pm}:\,M=\pm\delta, k_{\alpha}^{(1)}\leq k_{\alpha}\leq k_{\alpha}^{(2)}$ (yellow and blue planes in Fig.~\ref{fig:WeylNode}) and $\vb{B}(\vb{k},M)$ is the three-dimensional Berry curvature of $\ket{u_0^{(M)}}$ in the $(\vb{k},M)$ space. In this calculation, we used the fact that along $M=+\delta\ (-\delta)$ the band structure is gapped and we can take a gauge continuous for the whole surface $S^+\ (S^-)$.

We also note
\be
  \Delta\phi_{\beta}(k_{\alpha})\equiv 0 \qquad(\mathrm{mod}\ 2\pi)
\ee
as long as the system is gapped on the line $k_{\alpha}=const.$, because the occupied states are continuous and analytic across $M=0$, guaranteed by the gap. Here, the Berry phase is defined modulo $2\pi$ due to the gauge degree of freedom. Thus, we can write
\be
  \Delta\phi_{\beta}(k_{\alpha})=2\pi n(k_{\alpha}),\label{eq:Deltaphi}
\ee
where $n(k_{\alpha})$ is an integer dependent on $k_{\alpha}$. Here, because the gap is open at $M=+\delta$ and at $M=-\delta$, we take the gauge to be continuous in the entire two-dimensional Brillouin zone at $M=+\delta$ and likewise at $M=-\delta$.

\begin{figure}[t]
  \begin{center}
    \includegraphics[width=\columnwidth]{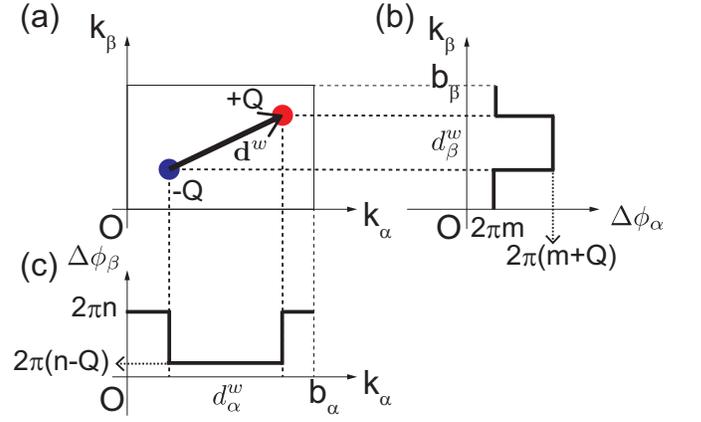}
    \caption{Schematic illustrations of (higher-order) Weyl nodes with monopole charges $\pm Q$. (a) Two Weyl nodes in a Brillouin zone. The black arrow indicates the displacement vector. (b) (c) The jumps of the differences of Berry phases in the $\beta(\alpha)$-direction plotted as a function of $k_{\alpha}(k_{\beta})$, respectively across $M=0$. They jump by $\pm 2\pi Q$ at the projections of the Weyl nodes to each axis.}
    \label{fig:General_Zak}
  \end{center}
\end{figure}

While the integer $n$ in Eq.~\eqref{eq:Deltaphi} can be chosen arbitrarily by gauge transformations, once we fix the gauge at a certain value of $k_{\alpha}$, the values of $\Delta\phi_{\beta}(k_{\alpha})$ is fixed for other values of $k_{\alpha}$ by Eq.~\eqref{eq:diffDeltaphi}. In Fig.~\ref{fig:General_Zak}(a), we illustrate this case with a pair of (higher-order) Weyl nodes with monopole charges $\pm Q$, and Fig.~\ref{fig:General_Zak}(c) shows the change of $\Delta \phi_{\beta}(k_{\alpha})$. One can study the difference of Berry phases in the other direction $\Delta\phi_{\alpha}(k_{\beta})$ in a similar way, and it is shown in Fig.~\ref{fig:General_Zak}(b).

Then, by integrating Berry phases, jumps of electric polarization in each direction are
\be
  \Delta P_{\alpha}&=\qty(-\frac{e}{2\pi}Qd_{\beta}^w-\frac{me}{2\pi}b_{\beta})\sin\theta,\\
  \Delta P_{\beta}&=\qty(\frac{e}{2\pi}Qd_{\alpha}^w-\frac{ne}{2\pi}b_{\alpha})\sin\theta,
\ee
where $m,n\in\mathbb{Z}$ and $\vb{d}^w=d_{\alpha}^w\vb{e}_{\alpha}+d_{\beta}^w\vb{e}_{\beta}$ $\qty(\vb{e}_{i}\coloneqq\vb{b}_i/\abs{\vb{b}_i},\,i=\alpha,\beta)$ represents the displacement vector pointing from the Weyl node with a negative monopole charge to that with a positive monopole charge in the reciprocal space.

Now we can introduce ``Weyl dipole" similarly to an electric dipole as
\be
 \vb{p}^w\coloneqq Q\vb{d}^w.
\ee
Then the polarization jump $\Delta \vb{P}$ can be written as
\be
  \Delta\vb{P}=\frac{e}{2\pi}\hat{z}\times\vb{p}^w+\frac{e}{\Omega}\vb{a},\label{eq:result:general_DeltaP}
\ee
where $\hat{z}=(0,0,1)^T$ and $\vb{a}\, (\coloneqq-m\vb{a}_{\alpha}-n\vb{a}_{\beta})$ is a lattice vector. Since $\vb{P}^+$ and $\vb{P}^-$ are determined in terms of modulo the quantum of polarization $\frac{e}{\Omega}\vb{a}_j\ (j=\alpha,\beta)$, $\Delta\vb{P}$ is also defined in terms of modulo $\frac{e}{\Omega}\vb{a}_j$ and the second term in Eq.~\eqref{eq:result:general_DeltaP} represents this degree of freedom.

Therefore, we can write the jump of electric polarization as
\be
  \Delta\vb{P}\equiv\frac{e}{2\pi}\hat{z}\times\vb{p}^w\qquad\qty(\mathrm{mod}\ \frac{e}{\Omega}\vb{a}_j).\label{eq:result:simple_DeltaP}
\ee
This is the main result of this paper. So far, we considered the simplest case with only one pair of Weyl nodes in the Brillouin zone; meanwhile, this formula \eqref{eq:result:simple_DeltaP} holds in general cases where more than two Weyl nodes or higher-order Weyl nodes (with $|Q|\geq2$) exist in the Brillouin zone. Still in such cases, one can define a total Weyl dipole $\vb{p}^w$ since the sum of monopole charges in the Brillouin zone must be zero. Then, the jump of the electric polarization is given by Eq.~\eqref{eq:result:simple_DeltaP}. We note that the Weyl dipole $\vb{p}^w$ is defined in terms of modulo $\vb{b}_i$ due to Brillouin zone periodicity; this does not affect the result \eqref{eq:result:simple_DeltaP} because it reduces to ``modulo-$\frac{e}{\Omega}\vb{a}_j$" ambiguity in Eq.~\eqref{eq:result:simple_DeltaP}.

Thus, we have shown that the jump of the polarization across $M=0$ (conducting limit) is given in a simple formula \eqref{eq:result:simple_DeltaP} in general. If we consider Weyl points protected by the inversion symmetry and assume that inversion-symmetry-breaking $M$ changes its sign under space inverion, we have $\vb{P}^+\equiv-\vb{P}^-\ \qty(\mathrm{mod}\ \frac{e}{\Omega}\vb{a}_j)$ and
\be
  \vb{P}^+&\equiv\frac{e}{4\pi} \hat{z}\times\vb{p}^w\qquad\qty(\mathrm{mod}\ \frac{e}{2\Omega}\vb{a}_j),\\
  \vb{P}^-&\equiv-\frac{e}{4\pi} \hat{z}\times\vb{p}^w\qquad\qty(\mathrm{mod}\ \frac{e}{2\Omega}\vb{a}_j)
\ee
holds in general multiband systems.

Finally, we confirm our general result by the example shown in the previous section. The monopole charges of the system are $Q=+1$ for the Weyl node $W_1$ and $Q=-1$ for $W_2$ when $-2<t_2/t_1<0$ (Fig.~\ref{fig:BN_lattice}(b)) and $Q=+1$ for $W_2$ and $Q=-1$ for $W_1$ when $0<t_2/t_1<2$ (Fig.~\ref{fig:BN_lattice}(d)). By using the positions of Weyl nodes Eqs.~\eqref{eq:W1} and \eqref{eq:W2}, the Weyl dipole is expressed as $\vb{p}^w = \qty(\frac{2}{\sqrt{3}a}\abs{\cos^{-1}\qty(\frac{-t_2}{2t_1})-\cos^{-1}\qty(\frac{t_2}{2t_1})},\mathrm{sgn}\qty(\frac{t_2}{t_1})\frac{2\pi}{3a},0)^T$. Therefore, the jump of polarization is
\be
  \Delta\vb{P} &\equiv \mqty(-\mathrm{sgn}\qty(\frac{t_2}{t_1})\frac{\sqrt{3}}{2}\\\frac{3}{2\pi}\abs{\cos^{-1}\qty(-\frac{t_2}{2t_1})-\cos^{-1}\qty(\frac{t_2}{2t_1})}\\0)P_0\nonumber\\
  &\equiv \mqty(0\\-\frac{3}{2}+\frac{3}{2\pi}\abs{\cos^{-1}\qty(-\frac{t_2}{2t_1})-\cos^{-1}\qty(\frac{t_2}{2t_1})}\\0)P_0\nonumber\\
  &\hspace{5cm}\qty(\mathrm{mod}\ \frac{e}{\Omega}\vb{a}_{1,2}).
\ee
This agrees with Eq.~\eqref{eq:AnalyticalP}.

\subsection{Candidate Materials}
\label{subsec:CandidateMaterials}

In this subsection, we discuss candidate materials of two-dimensional Weyl semimetals protected by some symmetry. To confirm the theory of this paper, here we also discuss external perturbations that break the symmetry and gap the Weyl nodes.

One option is to use $\mathcal{PT}$-symmetric Weyl semimetals such as graphene, AB-stacked bilayer graphene, graphynes~\cite{Malko2012,Zhao2020}, and Mg$_2$C monolayers~\cite{Wang2018}. Among them, in graphene, graphynes, and Mg$_2$C monolayers, the Weyl nodes are protected both by $\mathcal{PT}$ and $\mathcal{C}_{2z}\mathcal{T}$ symmetries. To open a gap at the Weyl nodes of these materials, we need to break both $\mathcal{PT}$ and $\mathcal{C}_{2z}\mathcal{T}$ symmetries. Therefore, if we use an electric field $\vb{E}$ for the symmetry-breaking term $M$, it should be in-plane, but in reality, the in-plane electric field simply induces a current, instead of opening a gap. Thus, we need to adopt other types of symmetry-breaking terms, such as a staggered potential as discussed in Sec.~\ref{sec:example}, as the symmetry-breaking term $M$.

On the other hand, AB-stacked bilayer graphene is known to host Weyl nodes with Berry phase $2\pi$ at $K$ and $K'$ points which can be gapped by applying an interlayer bias voltage $M$~\cite{McCann2006}. Then, it has Weyl nodes in the three-dimensional space $(k_x,k_y,M)$ with monopole charges $Q=\pm2$ and the situation is similar to that of Sec.~\ref{sec:example}. However, polarization in a two-dimensional sheet is prohibited by its three-fold rotational symmetry. Therefore, we need to displace Weyl nodes from $K$ and $K'$ points by breaking the rotational symmetry, for example by imposing a uniaxial strain to get a non-zero polarization. Then, our theory predicts a polarization jump by changing the interlayer bias voltage across zero. This idea of generating a polarization by interlayer bias voltage under the broken rotational symmetry in the bilayer graphene is similar to the one discussed in the context of nanoribbons of AB-stacked bilayer graphene~\cite{Okugawa2015}.

Other candidates can be Weyl semimetals without the inversion symmetry such as a strained Na$_2$O sheet, which hosts Weyl nodes on the $k_x$ axis protected both by a mirror symmetry $\mathcal{M}_y$ and a two-fold rotational symmetry $\mathcal{C}_{2x}$~\cite{Hua2020}, where the sheet lies in the $xy$ plane. For such systems, in order to open a gap only by an external electric field, one needs to break both of these symmetries. Only the electric field along the $y$-direction satisfies these conditions but it cannot be used since it just induces a current within the film. Instead, adding further in-plane strains in other directions is expected to solve this problem and make it possible to observe the polarization jump.

\section{Conclusion}
\label{sec:conclusion}

In summary, we have studied the behavior of the electric polarization in the limit where the two-dimensional system hosts symmetry protected Weyl nodes. We presented an example of a $\mathcal{PT}$-symmetric Weyl semimetal by a model on the hexagonal lattice with anisotropic hoppings and inversion-symmetry-breaking on-site potentials. Calculating the polarization using the modern theory of polarization, we found that in the inversion symmetric limit, the polarization approaches zero if the system is insulating but if the system becomes a Weyl semimetal in this limit, the polarization has a non-zero limit value. Furthermore, this limit value depends on whether the on-site potential approaches zero from the positive side or the negative side, and hence, there is a finite jump in polarization in the conducting limit. This jump of the polarization is expressed as a function of the distance between two Weyl nodes.

The analysis is extended to general two-dimensional systems with a symmetry-breaking parameter $M$, such that it is gapless at Weyl nodes at $M=0$ and insulating when $M\neq0$. Even if there are many bands below the Fermi energy, the jump of the polarization is determined by the states near the Weyl nodes, and it is universally written in terms of the ``Weyl dipole", which we introduced in this paper. The results indicate that the size of the jump does not depend on other details of the band structure. We discussed candidate materials for experimental observations of our proposal.

A similar phenomenon is reported in~\cite{Yu2020}, where a jump in the piezoelectric tensor (PET) across a topological phase transition at the gap closing is predicted in two-dimensional spinful systems. This proposal is distinct from that in the present paper. In our theory, the jump in the polarization is solely determined by the Weyl dipole, while the PET jump depends on several system parameters such as electron-strain coupling and Dirac velocities. Furthermore, the polarization jump can be seen both in spinless and spinful systems, and it is not necessarily accompanied by topological phase transition.

\begin{acknowledgements}
  This work is partly supported by Japan Society for the Promotion of Science (JSPS) KAKENHI Grants No. JP21K13865, No. JP22K18687, and No. JP22H00108.
\end{acknowledgements}

\bibliography{polarization.bib}

\begin{thebibliography}{28}%
\makeatletter
\providecommand \@ifxundefined [1]{%
 \@ifx{#1\undefined}
}%
\providecommand \@ifnum [1]{%
 \ifnum #1\expandafter \@firstoftwo
 \else \expandafter \@secondoftwo
 \fi
}%
\providecommand \@ifx [1]{%
 \ifx #1\expandafter \@firstoftwo
 \else \expandafter \@secondoftwo
 \fi
}%
\providecommand \natexlab [1]{#1}%
\providecommand \enquote  [1]{``#1''}%
\providecommand \bibnamefont  [1]{#1}%
\providecommand \bibfnamefont [1]{#1}%
\providecommand \citenamefont [1]{#1}%
\providecommand \href@noop [0]{\@secondoftwo}%
\providecommand \href [0]{\begingroup \@sanitize@url \@href}%
\providecommand \@href[1]{\@@startlink{#1}\@@href}%
\providecommand \@@href[1]{\endgroup#1\@@endlink}%
\providecommand \@sanitize@url [0]{\catcode `\\12\catcode `\$12\catcode
  `\&12\catcode `\#12\catcode `\^12\catcode `\_12\catcode `\%12\relax}%
\providecommand \@@startlink[1]{}%
\providecommand \@@endlink[0]{}%
\providecommand \url  [0]{\begingroup\@sanitize@url \@url }%
\providecommand \@url [1]{\endgroup\@href {#1}{\urlprefix }}%
\providecommand \urlprefix  [0]{URL }%
\providecommand \Eprint [0]{\href }%
\providecommand \doibase [0]{https://doi.org/}%
\providecommand \selectlanguage [0]{\@gobble}%
\providecommand \bibinfo  [0]{\@secondoftwo}%
\providecommand \bibfield  [0]{\@secondoftwo}%
\providecommand \translation [1]{[#1]}%
\providecommand \BibitemOpen [0]{}%
\providecommand \bibitemStop [0]{}%
\providecommand \bibitemNoStop [0]{.\EOS\space}%
\providecommand \EOS [0]{\spacefactor3000\relax}%
\providecommand \BibitemShut  [1]{\csname bibitem#1\endcsname}%
\let\auto@bib@innerbib\@empty
\bibitem [{\citenamefont {Resta}(1992)}]{Resta1992}%
  \BibitemOpen
  \bibfield  {author} {\bibinfo {author} {\bibfnamefont {R.}~\bibnamefont
  {Resta}},\ }\bibfield  {title} {\bibinfo {title} {{Theory of the electric
  polarization in crystals}},\ }\href
  {https://doi.org/10.1080/00150199208016065} {\bibfield  {journal} {\bibinfo
  {journal} {Ferroelectrics}\ }\textbf {\bibinfo {volume} {136}},\ \bibinfo
  {pages} {51} (\bibinfo {year} {1992})}\BibitemShut {NoStop}%
\bibitem [{\citenamefont {King-Smith}\ and\ \citenamefont
  {Vanderbilt}(1993)}]{King-Smith1993}%
  \BibitemOpen
  \bibfield  {author} {\bibinfo {author} {\bibfnamefont {R.~D.}\ \bibnamefont
  {King-Smith}}\ and\ \bibinfo {author} {\bibfnamefont {D.}~\bibnamefont
  {Vanderbilt}},\ }\bibfield  {title} {\bibinfo {title} {{Theory of
  polarization of crystalline solids}},\ }\href
  {https://doi.org/10.1103/PhysRevB.47.1651} {\bibfield  {journal} {\bibinfo
  {journal} {Phys. Rev. B}\ }\textbf {\bibinfo {volume} {47}},\ \bibinfo
  {pages} {1651} (\bibinfo {year} {1993})}\BibitemShut {NoStop}%
\bibitem [{\citenamefont {Vanderbilt}\ and\ \citenamefont
  {King-Smith}(1993)}]{Vanderbilt1993}%
  \BibitemOpen
  \bibfield  {author} {\bibinfo {author} {\bibfnamefont {D.}~\bibnamefont
  {Vanderbilt}}\ and\ \bibinfo {author} {\bibfnamefont {R.~D.}\ \bibnamefont
  {King-Smith}},\ }\bibfield  {title} {\bibinfo {title} {{Electric polarization
  as a bulk quantity and its relation to surface charge}},\ }\href
  {https://doi.org/10.1103/PhysRevB.48.4442} {\bibfield  {journal} {\bibinfo
  {journal} {Phys. Rev. B}\ }\textbf {\bibinfo {volume} {48}},\ \bibinfo
  {pages} {4442} (\bibinfo {year} {1993})}\BibitemShut {NoStop}%
\bibitem [{\citenamefont {Kane}\ and\ \citenamefont
  {Mele}(2005{\natexlab{a}})}]{Kane2005a}%
  \BibitemOpen
  \bibfield  {author} {\bibinfo {author} {\bibfnamefont {C.~L.}\ \bibnamefont
  {Kane}}\ and\ \bibinfo {author} {\bibfnamefont {E.~J.}\ \bibnamefont
  {Mele}},\ }\bibfield  {title} {\bibinfo {title} {{Z$_2$ Topological Order and
  the Quantum Spin Hall Effect}},\ }\href
  {https://doi.org/10.1103/PhysRevLett.95.146802} {\bibfield  {journal}
  {\bibinfo  {journal} {Phys. Rev. Lett.}\ }\textbf {\bibinfo {volume} {95}},\
  \bibinfo {pages} {146802} (\bibinfo {year} {2005}{\natexlab{a}})}\BibitemShut
  {NoStop}%
\bibitem [{\citenamefont {Kane}\ and\ \citenamefont
  {Mele}(2005{\natexlab{b}})}]{Kane2005b}%
  \BibitemOpen
  \bibfield  {author} {\bibinfo {author} {\bibfnamefont {C.~L.}\ \bibnamefont
  {Kane}}\ and\ \bibinfo {author} {\bibfnamefont {E.~J.}\ \bibnamefont
  {Mele}},\ }\bibfield  {title} {\bibinfo {title} {{Quantum Spin Hall Effect in
  Graphene}},\ }\href {https://doi.org/10.1103/PhysRevLett.95.226801}
  {\bibfield  {journal} {\bibinfo  {journal} {Phys. Rev. Lett.}\ }\textbf
  {\bibinfo {volume} {95}},\ \bibinfo {pages} {226801} (\bibinfo {year}
  {2005}{\natexlab{b}})}\BibitemShut {NoStop}%
\bibitem [{\citenamefont {Fu}\ \emph {et~al.}(2007)\citenamefont {Fu},
  \citenamefont {Kane},\ and\ \citenamefont {Mele}}]{Fu2007}%
  \BibitemOpen
  \bibfield  {author} {\bibinfo {author} {\bibfnamefont {L.}~\bibnamefont
  {Fu}}, \bibinfo {author} {\bibfnamefont {C.~L.}\ \bibnamefont {Kane}},\ and\
  \bibinfo {author} {\bibfnamefont {E.~J.}\ \bibnamefont {Mele}},\ }\bibfield
  {title} {\bibinfo {title} {{Topological Insulators in Three Dimensions}},\
  }\href {https://doi.org/10.1103/PhysRevLett.98.106803} {\bibfield  {journal}
  {\bibinfo  {journal} {Phys. Rev. Lett.}\ }\textbf {\bibinfo {volume} {98}},\
  \bibinfo {pages} {106803} (\bibinfo {year} {2007})}\BibitemShut {NoStop}%
\bibitem [{\citenamefont {K\"{o}nig}\ \emph {et~al.}(2007)\citenamefont
  {K\"{o}nig}, \citenamefont {Wiedmann}, \citenamefont {Br\"{u}ne},
  \citenamefont {Roth}, \citenamefont {Buhmann}, \citenamefont {Molenkamp},
  \citenamefont {Qi},\ and\ \citenamefont {Zhang}}]{Konig2010}%
  \BibitemOpen
  \bibfield  {author} {\bibinfo {author} {\bibfnamefont {M.}~\bibnamefont
  {K\"{o}nig}}, \bibinfo {author} {\bibfnamefont {S.}~\bibnamefont {Wiedmann}},
  \bibinfo {author} {\bibfnamefont {C.}~\bibnamefont {Br\"{u}ne}}, \bibinfo
  {author} {\bibfnamefont {A.}~\bibnamefont {Roth}}, \bibinfo {author}
  {\bibfnamefont {H.}~\bibnamefont {Buhmann}}, \bibinfo {author} {\bibfnamefont
  {L.~W.}\ \bibnamefont {Molenkamp}}, \bibinfo {author} {\bibfnamefont {X.-L.}\
  \bibnamefont {Qi}},\ and\ \bibinfo {author} {\bibfnamefont {S.-C.}\
  \bibnamefont {Zhang}},\ }\bibfield  {title} {\bibinfo {title} {{Quantum Spin
  Hall Insulator State in HgTe Quantum Wells}},\ }\href
  {https://doi.org/10.1126/science.1148047} {\bibfield  {journal} {\bibinfo
  {journal} {Science}\ }\textbf {\bibinfo {volume} {318}},\ \bibinfo {pages}
  {766} (\bibinfo {year} {2007})}\BibitemShut {NoStop}%
\bibitem [{\citenamefont {Bernevig}\ \emph {et~al.}(2006)\citenamefont
  {Bernevig}, \citenamefont {Hughes},\ and\ \citenamefont
  {Zhang}}]{Bernevig2006}%
  \BibitemOpen
  \bibfield  {author} {\bibinfo {author} {\bibfnamefont {B.~A.}\ \bibnamefont
  {Bernevig}}, \bibinfo {author} {\bibfnamefont {T.~L.}\ \bibnamefont
  {Hughes}},\ and\ \bibinfo {author} {\bibfnamefont {S.-C.}\ \bibnamefont
  {Zhang}},\ }\bibfield  {title} {\bibinfo {title} {{Quantum Spin Hall Effect
  and Topological Phase Transition in HgTe Quantum Wells}},\ }\href
  {https://doi.org/10.1126/science.1133734} {\bibfield  {journal} {\bibinfo
  {journal} {Science}\ }\textbf {\bibinfo {volume} {314}},\ \bibinfo {pages}
  {1757} (\bibinfo {year} {2006})}\BibitemShut {NoStop}%
\bibitem [{\citenamefont {Wan}\ \emph {et~al.}(2011)\citenamefont {Wan},
  \citenamefont {Turner}, \citenamefont {Vishwanath},\ and\ \citenamefont
  {Savrasov}}]{Wan2011}%
  \BibitemOpen
  \bibfield  {author} {\bibinfo {author} {\bibfnamefont {X.}~\bibnamefont
  {Wan}}, \bibinfo {author} {\bibfnamefont {A.~M.}\ \bibnamefont {Turner}},
  \bibinfo {author} {\bibfnamefont {A.}~\bibnamefont {Vishwanath}},\ and\
  \bibinfo {author} {\bibfnamefont {S.~Y.}\ \bibnamefont {Savrasov}},\
  }\bibfield  {title} {\bibinfo {title} {{Topological semimetal and Fermi-arc
  surface states in the electronic structure of pyrochlore iridates}},\ }\href
  {https://doi.org/10.1103/PhysRevB.83.205101} {\bibfield  {journal} {\bibinfo
  {journal} {Phys. Rev. B}\ }\textbf {\bibinfo {volume} {83}},\ \bibinfo
  {pages} {205101} (\bibinfo {year} {2011})}\BibitemShut {NoStop}%
\bibitem [{\citenamefont {Huang}\ \emph {et~al.}(2015)\citenamefont {Huang},
  \citenamefont {Xu}, \citenamefont {Belopolski}, \citenamefont {Lee},
  \citenamefont {Chang}, \citenamefont {Wang}, \citenamefont {Alidoust},
  \citenamefont {Bian}, \citenamefont {Neupane}, \citenamefont {Zhang},
  \citenamefont {Jia}, \citenamefont {Bansil}, \citenamefont {Lin},\ and\
  \citenamefont {Hasan}}]{Huang2015}%
  \BibitemOpen
  \bibfield  {author} {\bibinfo {author} {\bibfnamefont {S.-M.}\ \bibnamefont
  {Huang}}, \bibinfo {author} {\bibfnamefont {S.-Y.}\ \bibnamefont {Xu}},
  \bibinfo {author} {\bibfnamefont {I.}~\bibnamefont {Belopolski}}, \bibinfo
  {author} {\bibfnamefont {C.-C.}\ \bibnamefont {Lee}}, \bibinfo {author}
  {\bibfnamefont {G.}~\bibnamefont {Chang}}, \bibinfo {author} {\bibfnamefont
  {B.}~\bibnamefont {Wang}}, \bibinfo {author} {\bibfnamefont {N.}~\bibnamefont
  {Alidoust}}, \bibinfo {author} {\bibfnamefont {G.}~\bibnamefont {Bian}},
  \bibinfo {author} {\bibfnamefont {M.}~\bibnamefont {Neupane}}, \bibinfo
  {author} {\bibfnamefont {C.}~\bibnamefont {Zhang}}, \bibinfo {author}
  {\bibfnamefont {S.}~\bibnamefont {Jia}}, \bibinfo {author} {\bibfnamefont
  {A.}~\bibnamefont {Bansil}}, \bibinfo {author} {\bibfnamefont
  {H.}~\bibnamefont {Lin}},\ and\ \bibinfo {author} {\bibfnamefont {M.~Z.}\
  \bibnamefont {Hasan}},\ }\bibfield  {title} {\bibinfo {title} {{A Weyl
  Fermion semimetal with surface Fermi arcs in the transition metal
  monopnictide TaAs class}},\ }\href {https://doi.org/10.1038/ncomms8373}
  {\bibfield  {journal} {\bibinfo  {journal} {Nat. Commun.}\ }\textbf {\bibinfo
  {volume} {6}},\ \bibinfo {pages} {7373} (\bibinfo {year} {2015})}\BibitemShut
  {NoStop}%
\bibitem [{\citenamefont {Lv}\ \emph {et~al.}(2015)\citenamefont {Lv},
  \citenamefont {Weng}, \citenamefont {Fu}, \citenamefont {Wang}, \citenamefont
  {Miao}, \citenamefont {Ma}, \citenamefont {Richard}, \citenamefont {Huang},
  \citenamefont {Zhao}, \citenamefont {Chen}, \citenamefont {Fang},
  \citenamefont {Dai}, \citenamefont {Qian},\ and\ \citenamefont
  {Ding}}]{Lv2015}%
  \BibitemOpen
  \bibfield  {author} {\bibinfo {author} {\bibfnamefont {B.~Q.}\ \bibnamefont
  {Lv}}, \bibinfo {author} {\bibfnamefont {H.~M.}\ \bibnamefont {Weng}},
  \bibinfo {author} {\bibfnamefont {B.~B.}\ \bibnamefont {Fu}}, \bibinfo
  {author} {\bibfnamefont {X.~P.}\ \bibnamefont {Wang}}, \bibinfo {author}
  {\bibfnamefont {H.}~\bibnamefont {Miao}}, \bibinfo {author} {\bibfnamefont
  {J.}~\bibnamefont {Ma}}, \bibinfo {author} {\bibfnamefont {P.}~\bibnamefont
  {Richard}}, \bibinfo {author} {\bibfnamefont {X.~C.}\ \bibnamefont {Huang}},
  \bibinfo {author} {\bibfnamefont {L.~X.}\ \bibnamefont {Zhao}}, \bibinfo
  {author} {\bibfnamefont {G.~F.}\ \bibnamefont {Chen}}, \bibinfo {author}
  {\bibfnamefont {Z.}~\bibnamefont {Fang}}, \bibinfo {author} {\bibfnamefont
  {X.}~\bibnamefont {Dai}}, \bibinfo {author} {\bibfnamefont {T.}~\bibnamefont
  {Qian}},\ and\ \bibinfo {author} {\bibfnamefont {H.}~\bibnamefont {Ding}},\
  }\bibfield  {title} {\bibinfo {title} {{Experimental Discovery of Weyl
  Semimetal TaAs}},\ }\href {https://doi.org/10.1103/PhysRevX.5.031013}
  {\bibfield  {journal} {\bibinfo  {journal} {Phys. Rev. X}\ }\textbf {\bibinfo
  {volume} {5}},\ \bibinfo {pages} {031013} (\bibinfo {year}
  {2015})}\BibitemShut {NoStop}%
\bibitem [{\citenamefont {Xu}\ \emph {et~al.}(2015)\citenamefont {Xu},
  \citenamefont {Belopolski}, \citenamefont {Alidoust}, \citenamefont
  {Neupane}, \citenamefont {Bian}, \citenamefont {Zhang}, \citenamefont
  {Sankar}, \citenamefont {Chang}, \citenamefont {Yuan}, \citenamefont {Lee},
  \citenamefont {Huang}, \citenamefont {Zheng}, \citenamefont {Ma},
  \citenamefont {Sanchez}, \citenamefont {Wang}, \citenamefont {Bansil},
  \citenamefont {Chou}, \citenamefont {Shibayev}, \citenamefont {Lin},
  \citenamefont {Jia},\ and\ \citenamefont {Hasan}}]{Xu2015}%
  \BibitemOpen
  \bibfield  {author} {\bibinfo {author} {\bibfnamefont {S.-Y.}\ \bibnamefont
  {Xu}}, \bibinfo {author} {\bibfnamefont {I.}~\bibnamefont {Belopolski}},
  \bibinfo {author} {\bibfnamefont {N.}~\bibnamefont {Alidoust}}, \bibinfo
  {author} {\bibfnamefont {M.}~\bibnamefont {Neupane}}, \bibinfo {author}
  {\bibfnamefont {G.}~\bibnamefont {Bian}}, \bibinfo {author} {\bibfnamefont
  {C.}~\bibnamefont {Zhang}}, \bibinfo {author} {\bibfnamefont
  {R.}~\bibnamefont {Sankar}}, \bibinfo {author} {\bibfnamefont
  {G.}~\bibnamefont {Chang}}, \bibinfo {author} {\bibfnamefont
  {Z.}~\bibnamefont {Yuan}}, \bibinfo {author} {\bibfnamefont {C.-C.}\
  \bibnamefont {Lee}}, \bibinfo {author} {\bibfnamefont {S.-M.}\ \bibnamefont
  {Huang}}, \bibinfo {author} {\bibfnamefont {H.}~\bibnamefont {Zheng}},
  \bibinfo {author} {\bibfnamefont {J.}~\bibnamefont {Ma}}, \bibinfo {author}
  {\bibfnamefont {D.~S.}\ \bibnamefont {Sanchez}}, \bibinfo {author}
  {\bibfnamefont {B.}~\bibnamefont {Wang}}, \bibinfo {author} {\bibfnamefont
  {A.}~\bibnamefont {Bansil}}, \bibinfo {author} {\bibfnamefont
  {F.}~\bibnamefont {Chou}}, \bibinfo {author} {\bibfnamefont {P.~P.}\
  \bibnamefont {Shibayev}}, \bibinfo {author} {\bibfnamefont {H.}~\bibnamefont
  {Lin}}, \bibinfo {author} {\bibfnamefont {S.}~\bibnamefont {Jia}},\ and\
  \bibinfo {author} {\bibfnamefont {M.~Z.}\ \bibnamefont {Hasan}},\ }\bibfield
  {title} {\bibinfo {title} {{Discovery of a Weyl fermion semimetal and
  topological Fermi arcs}},\ }\href {https://doi.org/10.1126/science.aaa9297}
  {\bibfield  {journal} {\bibinfo  {journal} {Science}\ }\textbf {\bibinfo
  {volume} {349}},\ \bibinfo {pages} {613} (\bibinfo {year}
  {2015})}\BibitemShut {NoStop}%
\bibitem [{\citenamefont {Wang}\ \emph {et~al.}(2012)\citenamefont {Wang},
  \citenamefont {Sun}, \citenamefont {Chen}, \citenamefont {Franchini},
  \citenamefont {Xu}, \citenamefont {Weng}, \citenamefont {Dai},\ and\
  \citenamefont {Fang}}]{Wang2012}%
  \BibitemOpen
  \bibfield  {author} {\bibinfo {author} {\bibfnamefont {Z.}~\bibnamefont
  {Wang}}, \bibinfo {author} {\bibfnamefont {Y.}~\bibnamefont {Sun}}, \bibinfo
  {author} {\bibfnamefont {X.-Q.}\ \bibnamefont {Chen}}, \bibinfo {author}
  {\bibfnamefont {C.}~\bibnamefont {Franchini}}, \bibinfo {author}
  {\bibfnamefont {G.}~\bibnamefont {Xu}}, \bibinfo {author} {\bibfnamefont
  {H.}~\bibnamefont {Weng}}, \bibinfo {author} {\bibfnamefont {X.}~\bibnamefont
  {Dai}},\ and\ \bibinfo {author} {\bibfnamefont {Z.}~\bibnamefont {Fang}},\
  }\bibfield  {title} {\bibinfo {title} {{Dirac semimetal and topological phase
  transitions in ${A}_{3}\mathrm{Bi}$
  (${A}=\mathrm{Na},\mathrm{K},\mathrm{Rb}$)}},\ }\href
  {https://doi.org/10.1103/PhysRevB.85.195320} {\bibfield  {journal} {\bibinfo
  {journal} {Phys. Rev. B}\ }\textbf {\bibinfo {volume} {85}},\ \bibinfo
  {pages} {195320} (\bibinfo {year} {2012})}\BibitemShut {NoStop}%
\bibitem [{\citenamefont {Young}\ \emph {et~al.}(2012)\citenamefont {Young},
  \citenamefont {Zaheer}, \citenamefont {Teo}, \citenamefont {Kane},
  \citenamefont {Mele},\ and\ \citenamefont {Rappe}}]{Young2012}%
  \BibitemOpen
  \bibfield  {author} {\bibinfo {author} {\bibfnamefont {S.~M.}\ \bibnamefont
  {Young}}, \bibinfo {author} {\bibfnamefont {S.}~\bibnamefont {Zaheer}},
  \bibinfo {author} {\bibfnamefont {J.~C.~Y.}\ \bibnamefont {Teo}}, \bibinfo
  {author} {\bibfnamefont {C.~L.}\ \bibnamefont {Kane}}, \bibinfo {author}
  {\bibfnamefont {E.~J.}\ \bibnamefont {Mele}},\ and\ \bibinfo {author}
  {\bibfnamefont {A.~M.}\ \bibnamefont {Rappe}},\ }\bibfield  {title} {\bibinfo
  {title} {{Dirac Semimetal in Three Dimensions}},\ }\href
  {https://doi.org/10.1103/PhysRevLett.108.140405} {\bibfield  {journal}
  {\bibinfo  {journal} {Phys. Rev. Lett.}\ }\textbf {\bibinfo {volume} {108}},\
  \bibinfo {pages} {140405} (\bibinfo {year} {2012})}\BibitemShut {NoStop}%
\bibitem [{\citenamefont {Young}\ and\ \citenamefont {Kane}(2015)}]{Young2015}%
  \BibitemOpen
  \bibfield  {author} {\bibinfo {author} {\bibfnamefont {S.~M.}\ \bibnamefont
  {Young}}\ and\ \bibinfo {author} {\bibfnamefont {C.~L.}\ \bibnamefont
  {Kane}},\ }\bibfield  {title} {\bibinfo {title} {{Dirac Semimetals in Two
  Dimensions}},\ }\href {https://doi.org/10.1103/PhysRevLett.115.126803}
  {\bibfield  {journal} {\bibinfo  {journal} {Phys. Rev. Lett.}\ }\textbf
  {\bibinfo {volume} {115}},\ \bibinfo {pages} {126803} (\bibinfo {year}
  {2015})}\BibitemShut {NoStop}%
\bibitem [{\citenamefont {Vazifeh}\ and\ \citenamefont
  {Franz}(2013)}]{Vazifeh2013}%
  \BibitemOpen
  \bibfield  {author} {\bibinfo {author} {\bibfnamefont {M.~M.}\ \bibnamefont
  {Vazifeh}}\ and\ \bibinfo {author} {\bibfnamefont {M.}~\bibnamefont
  {Franz}},\ }\bibfield  {title} {\bibinfo {title} {{Electromagnetic Response
  of Weyl Semimetals}},\ }\href
  {https://doi.org/10.1103/PhysRevLett.111.027201} {\bibfield  {journal}
  {\bibinfo  {journal} {Phys. Rev. Lett.}\ }\textbf {\bibinfo {volume} {111}},\
  \bibinfo {pages} {027201} (\bibinfo {year} {2013})}\BibitemShut {NoStop}%
\bibitem [{\citenamefont {Chan}\ \emph {et~al.}(2017)\citenamefont {Chan},
  \citenamefont {Lindner}, \citenamefont {Refael},\ and\ \citenamefont
  {Lee}}]{Chan2017}%
  \BibitemOpen
  \bibfield  {author} {\bibinfo {author} {\bibfnamefont {C.-K.}\ \bibnamefont
  {Chan}}, \bibinfo {author} {\bibfnamefont {N.~H.}\ \bibnamefont {Lindner}},
  \bibinfo {author} {\bibfnamefont {G.}~\bibnamefont {Refael}},\ and\ \bibinfo
  {author} {\bibfnamefont {P.~A.}\ \bibnamefont {Lee}},\ }\bibfield  {title}
  {\bibinfo {title} {{Photocurrents in Weyl semimetals}},\ }\href
  {https://doi.org/10.1103/PhysRevB.95.041104} {\bibfield  {journal} {\bibinfo
  {journal} {Phys. Rev. B}\ }\textbf {\bibinfo {volume} {95}},\ \bibinfo
  {pages} {041104} (\bibinfo {year} {2017})}\BibitemShut {NoStop}%
\bibitem [{\citenamefont {Gorbar}\ \emph {et~al.}(2018)\citenamefont {Gorbar},
  \citenamefont {Miransky}, \citenamefont {Shovkovy},\ and\ \citenamefont
  {Sukhachov}}]{Gorbar2018}%
  \BibitemOpen
  \bibfield  {author} {\bibinfo {author} {\bibfnamefont {E.~V.}\ \bibnamefont
  {Gorbar}}, \bibinfo {author} {\bibfnamefont {V.~A.}\ \bibnamefont
  {Miransky}}, \bibinfo {author} {\bibfnamefont {I.~A.}\ \bibnamefont
  {Shovkovy}},\ and\ \bibinfo {author} {\bibfnamefont {P.~O.}\ \bibnamefont
  {Sukhachov}},\ }\bibfield  {title} {\bibinfo {title} {{Anomalous transport
  properties of Dirac and Weyl semimetals (Review Article)}},\ }\href
  {https://doi.org/10.1063/1.5037551} {\bibfield  {journal} {\bibinfo
  {journal} {Low Temp. Phys.}\ }\textbf {\bibinfo {volume} {44}},\ \bibinfo
  {pages} {487} (\bibinfo {year} {2018})}\BibitemShut {NoStop}%
\bibitem [{\citenamefont {Mahon}\ \emph {et~al.}(2019)\citenamefont {Mahon},
  \citenamefont {Muniz},\ and\ \citenamefont {Sipe}}]{Mahon2019}%
  \BibitemOpen
  \bibfield  {author} {\bibinfo {author} {\bibfnamefont {P.~T.}\ \bibnamefont
  {Mahon}}, \bibinfo {author} {\bibfnamefont {R.~A.}\ \bibnamefont {Muniz}},\
  and\ \bibinfo {author} {\bibfnamefont {J.~E.}\ \bibnamefont {Sipe}},\
  }\bibfield  {title} {\bibinfo {title} {{Microscopic polarization and
  magnetization fields in extended systems}},\ }\href
  {https://doi.org/10.1103/PhysRevB.99.235140} {\bibfield  {journal} {\bibinfo
  {journal} {Phys. Rev. B}\ }\textbf {\bibinfo {volume} {99}},\ \bibinfo
  {pages} {235140} (\bibinfo {year} {2019})}\BibitemShut {NoStop}%
\bibitem [{\citenamefont {Mahon}\ and\ \citenamefont {Sipe}()}]{Mahon2021}%
  \BibitemOpen
  \bibfield  {author} {\bibinfo {author} {\bibfnamefont {P.~T.}\ \bibnamefont
  {Mahon}}\ and\ \bibinfo {author} {\bibfnamefont {J.~E.}\ \bibnamefont
  {Sipe}},\ }\bibfield  {title} {\bibinfo {title} {{Electric polarization and
  magnetization in metals}},\ }\Eprint {https://arxiv.org/abs/2111.03032}
  {arXiv:2111.03032} \BibitemShut {NoStop}%
\bibitem [{\citenamefont {Ohtomo}\ and\ \citenamefont
  {Hwang}(2004)}]{Ohtomo2004}%
  \BibitemOpen
  \bibfield  {author} {\bibinfo {author} {\bibfnamefont {A.}~\bibnamefont
  {Ohtomo}}\ and\ \bibinfo {author} {\bibfnamefont {H.~Y.}\ \bibnamefont
  {Hwang}},\ }\bibfield  {title} {\bibinfo {title} {{A high-mobility electron
  gas at the $\mathrm{LaAlO}_3/\mathrm{SrTiO}_3$ heterointerface}},\ }\href
  {https://doi.org/10.1038/nature02308} {\bibfield  {journal} {\bibinfo
  {journal} {Nature}\ }\textbf {\bibinfo {volume} {427}},\ \bibinfo {pages}
  {423} (\bibinfo {year} {2004})}\BibitemShut {NoStop}%
\bibitem [{\citenamefont {Malko}\ \emph {et~al.}(2012)\citenamefont {Malko},
  \citenamefont {Neiss}, \citenamefont {Vi\~nes},\ and\ \citenamefont
  {G\"orling}}]{Malko2012}%
  \BibitemOpen
  \bibfield  {author} {\bibinfo {author} {\bibfnamefont {D.}~\bibnamefont
  {Malko}}, \bibinfo {author} {\bibfnamefont {C.}~\bibnamefont {Neiss}},
  \bibinfo {author} {\bibfnamefont {F.}~\bibnamefont {Vi\~nes}},\ and\ \bibinfo
  {author} {\bibfnamefont {A.}~\bibnamefont {G\"orling}},\ }\bibfield  {title}
  {\bibinfo {title} {{Competition for Graphene: Graphynes with
  Direction-Dependent Dirac Cones}},\ }\href
  {https://doi.org/10.1103/PhysRevLett.108.086804} {\bibfield  {journal}
  {\bibinfo  {journal} {Phys. Rev. Lett.}\ }\textbf {\bibinfo {volume} {108}},\
  \bibinfo {pages} {086804} (\bibinfo {year} {2012})}\BibitemShut {NoStop}%
\bibitem [{\citenamefont {Zhao}\ \emph {et~al.}(2020)\citenamefont {Zhao},
  \citenamefont {Cui}, \citenamefont {Cai}, \citenamefont {Guo}, \citenamefont
  {Cui}, \citenamefont {Song},\ and\ \citenamefont {Liu}}]{Zhao2020}%
  \BibitemOpen
  \bibfield  {author} {\bibinfo {author} {\bibfnamefont {D.}~\bibnamefont
  {Zhao}}, \bibinfo {author} {\bibfnamefont {L.}~\bibnamefont {Cui}}, \bibinfo
  {author} {\bibfnamefont {J.}~\bibnamefont {Cai}}, \bibinfo {author}
  {\bibfnamefont {Y.}~\bibnamefont {Guo}}, \bibinfo {author} {\bibfnamefont
  {X.}~\bibnamefont {Cui}}, \bibinfo {author} {\bibfnamefont {T.}~\bibnamefont
  {Song}},\ and\ \bibinfo {author} {\bibfnamefont {Z.}~\bibnamefont {Liu}},\
  }\bibfield  {title} {\bibinfo {title} {{Palgraphyne: A Promising 2D Carbon
  Dirac Semimetal with Strong Mechanical and Electronic Anisotropy}},\ }\href
  {https://doi.org/10.1002/pssr.201900670} {\bibfield  {journal} {\bibinfo
  {journal} {physica status solidi (RRL) – Rapid Research Letters}\ }\textbf
  {\bibinfo {volume} {14}},\ \bibinfo {pages} {1900670} (\bibinfo {year}
  {2020})}\BibitemShut {NoStop}%
\bibitem [{\citenamefont {Wang}\ \emph {et~al.}(2018)\citenamefont {Wang},
  \citenamefont {Liu}, \citenamefont {Yu}, \citenamefont {Sheng}, \citenamefont
  {Zhu}, \citenamefont {Guan},\ and\ \citenamefont {Yang}}]{Wang2018}%
  \BibitemOpen
  \bibfield  {author} {\bibinfo {author} {\bibfnamefont {S.-S.}\ \bibnamefont
  {Wang}}, \bibinfo {author} {\bibfnamefont {Y.}~\bibnamefont {Liu}}, \bibinfo
  {author} {\bibfnamefont {Z.-M.}\ \bibnamefont {Yu}}, \bibinfo {author}
  {\bibfnamefont {X.-L.}\ \bibnamefont {Sheng}}, \bibinfo {author}
  {\bibfnamefont {L.}~\bibnamefont {Zhu}}, \bibinfo {author} {\bibfnamefont
  {S.}~\bibnamefont {Guan}},\ and\ \bibinfo {author} {\bibfnamefont {S.~A.}\
  \bibnamefont {Yang}},\ }\bibfield  {title} {\bibinfo {title} {{Monolayer
  ${\mathrm{Mg}}_{2}\mathrm{C}$: Negative Poisson's ratio and unconventional
  two-dimensional emergent fermions}},\ }\href
  {https://doi.org/10.1103/PhysRevMaterials.2.104003} {\bibfield  {journal}
  {\bibinfo  {journal} {Phys. Rev. Materials}\ }\textbf {\bibinfo {volume}
  {2}},\ \bibinfo {pages} {104003} (\bibinfo {year} {2018})}\BibitemShut
  {NoStop}%
\bibitem [{\citenamefont {McCann}\ and\ \citenamefont
  {Fal'ko}(2006)}]{McCann2006}%
  \BibitemOpen
  \bibfield  {author} {\bibinfo {author} {\bibfnamefont {E.}~\bibnamefont
  {McCann}}\ and\ \bibinfo {author} {\bibfnamefont {V.~I.}\ \bibnamefont
  {Fal'ko}},\ }\bibfield  {title} {\bibinfo {title} {{Landau-Level Degeneracy
  and Quantum Hall Effect in a Graphite Bilayer}},\ }\href
  {https://doi.org/10.1103/PhysRevLett.96.086805} {\bibfield  {journal}
  {\bibinfo  {journal} {Phys. Rev. Lett.}\ }\textbf {\bibinfo {volume} {96}},\
  \bibinfo {pages} {086805} (\bibinfo {year} {2006})}\BibitemShut {NoStop}%
\bibitem [{\citenamefont {Okugawa}\ \emph {et~al.}(2015)\citenamefont
  {Okugawa}, \citenamefont {Tanaka}, \citenamefont {Koretsune}, \citenamefont
  {Saito},\ and\ \citenamefont {Murakami}}]{Okugawa2015}%
  \BibitemOpen
  \bibfield  {author} {\bibinfo {author} {\bibfnamefont {R.}~\bibnamefont
  {Okugawa}}, \bibinfo {author} {\bibfnamefont {J.}~\bibnamefont {Tanaka}},
  \bibinfo {author} {\bibfnamefont {T.}~\bibnamefont {Koretsune}}, \bibinfo
  {author} {\bibfnamefont {S.}~\bibnamefont {Saito}},\ and\ \bibinfo {author}
  {\bibfnamefont {S.}~\bibnamefont {Murakami}},\ }\bibfield  {title} {\bibinfo
  {title} {{In-Plane Electric Polarization of Bilayer Graphene Nanoribbons
  Induced by an Interlayer Bias Voltage}},\ }\href
  {https://doi.org/10.1103/PhysRevLett.115.156601} {\bibfield  {journal}
  {\bibinfo  {journal} {Phys. Rev. Lett.}\ }\textbf {\bibinfo {volume} {115}},\
  \bibinfo {pages} {156601} (\bibinfo {year} {2015})}\BibitemShut {NoStop}%
\bibitem [{\citenamefont {Hua}\ \emph {et~al.}(2020)\citenamefont {Hua},
  \citenamefont {Li}, \citenamefont {Xu}, \citenamefont {Zheng}, \citenamefont
  {Yang},\ and\ \citenamefont {Lu}}]{Hua2020}%
  \BibitemOpen
  \bibfield  {author} {\bibinfo {author} {\bibfnamefont {C.}~\bibnamefont
  {Hua}}, \bibinfo {author} {\bibfnamefont {S.}~\bibnamefont {Li}}, \bibinfo
  {author} {\bibfnamefont {Z.-A.}\ \bibnamefont {Xu}}, \bibinfo {author}
  {\bibfnamefont {Y.}~\bibnamefont {Zheng}}, \bibinfo {author} {\bibfnamefont
  {S.~A.}\ \bibnamefont {Yang}},\ and\ \bibinfo {author} {\bibfnamefont
  {Y.}~\bibnamefont {Lu}},\ }\bibfield  {title} {\bibinfo {title} {{Tunable
  Topological Energy Bands in 2D Dialkali-Metal Monoxides}},\ }\href
  {https://doi.org/10.1002/advs.201901939} {\bibfield  {journal} {\bibinfo
  {journal} {Advanced Science}\ }\textbf {\bibinfo {volume} {7}},\ \bibinfo
  {pages} {1901939} (\bibinfo {year} {2020})}\BibitemShut {NoStop}%
\bibitem [{\citenamefont {Yu}\ and\ \citenamefont {Liu}(2020)}]{Yu2020}%
  \BibitemOpen
  \bibfield  {author} {\bibinfo {author} {\bibfnamefont {J.}~\bibnamefont
  {Yu}}\ and\ \bibinfo {author} {\bibfnamefont {C.-X.}\ \bibnamefont {Liu}},\
  }\bibfield  {title} {\bibinfo {title} {{Piezoelectricity and Topological
  Quantum Phase Transitions in Two-Dimensional Spin-Orbit Coupled Crystals with
  Time-Reversal Symmetry}},\ }\href
  {https://doi.org/10.1038/s41467-020-16058-2} {\bibfield  {journal} {\bibinfo
  {journal} {Nat. Commun.}\ }\textbf {\bibinfo {volume} {11}},\ \bibinfo
  {pages} {2290} (\bibinfo {year} {2020})}\BibitemShut {NoStop}%
\end{thebibliography}%

\end{document}